\documentclass[runningheads]{llncs}
\usepackage[T1]{fontenc}
\usepackage{graphicx}%
\usepackage{multirow}%
\usepackage{amsmath,amssymb,amsfonts}%
\usepackage{mathrsfs}%
\usepackage[title]{appendix}%
\usepackage{xcolor}%
\usepackage{textcomp}%
\usepackage{manyfoot}%
\usepackage{booktabs}%
\usepackage{tabularx}
\usepackage{subfigure}

\usepackage{listings}%
\usepackage[ruled,vlined]{algorithm2e}
\usepackage[table]{xcolor}

\newcommand{\overallbest}[1]{\cellcolor{gray!15}\textbf{#1}}

\newcommand{\smallsection}[1]{\vspace{1mm}\noindent\smash{\textbf{#1.}}}

\let\llncssubparagraph\subparagraph
\let\subparagraph\paragraph
\usepackage{titlesec}
\let\subparagraph\llncssubparagraph
\titlespacing{\section}{0pt}{*1}{*1}
\titlespacing{\subsection}{0pt}{*1}{*0.5}

\begin{document}
%
\title{
Automatic In-Domain Exemplar Construction and LLM-Based Refinement of Multi-LLM Expansions for Query Expansion\thanks{Preprint. This paper is under consideration at Pattern Recognition Letters.}
}
\titlerunning{LLM-Based Refinement of Multi-LLM Expansions for QE}
%
\author{Minghan Li\inst{1}\thanks{Corresponding author.} \and
Ercong Nie\inst{2,3} \and
Siqi Zhao\inst{1} \and
Tongna Chen\inst{1} \and
Huiping Huang\inst{4} \and
Guodong Zhou\inst{1}
}
\authorrunning{Minghan Li et al.}
%
\institute{School of Computer Science and Technology, Soochow University, Suzhou, China \\
\email{\{mhli,gdzhou\}@suda.edu.cn, \{sqzhaosqzhao,tnchentnchen\}@stu.suda.edu.cn} \\
\and
Center for Information and Language Processing (CIS), LMU Munich, Munich, Germany
\and
Munich Center for Machine Learning (MCML), Munich, Germany\\
\email{nie@cis.lmu.de}
\and
Chalmers University of Technology, Gothenburg, Sweden\\
\email{huiping@chalmers.se}
}

\maketitle              
\begin{abstract}
Query expansion with large language models is promising but often relies on hand-crafted prompts, manually chosen exemplars, or a single LLM, making it non-scalable and sensitive to domain shift. We present an automated, domain-adaptive QE framework that builds in-domain exemplar pools by harvesting pseudo-relevant passages using a BM25–MonoT5 pipeline. A training-free cluster-based strategy selects diverse demonstrations, yielding strong and stable in-context QE without supervision. To further exploit model complementarity, we introduce a two-LLM ensemble in which two heterogeneous LLMs independently generate expansions and a refinement LLM consolidates them into one coherent expansion. Across TREC DL20, DBPedia, and SciFact, the refined ensemble delivers consistent and statistically significant gains over BM25, Rocchio, zero-shot, and fixed few-shot baselines. The framework offers a reproducible testbed for exemplar selection and multi-LLM generation, and a practical, label-free solution for real-world QE.

\end{abstract}

\section{Introduction}

Query expansion (QE) has long been used to mitigate the vocabulary mismatch
between user queries and relevant documents \cite{wang2023query2doc}. 
Classical pseudo-relevance feedback (PRF) methods such as Rocchio 
\cite{cao2008selecting,miao2012proximity,liu2022simple} and RM3 
\cite{abdul2004umass} expand queries using terms harvested from an initial
BM25 retrieval.  
While effective, PRF depends heavily on the quality of the first-stage
ranker and cannot inject semantic knowledge beyond the corpus.

Large language models (LLMs) \cite{kalyan2024survey} introduce a new
paradigm by generating semantically rich reformulations of user queries.
In-context learning (ICL) further enables few-shot QE by prompting the
LLM with several example query–passage pairs without any parameter updates
\cite{dong2024survey}.  
However, ICL is notoriously sensitive to the choice and ordering of
examples \cite{NEURIPS2020_1457c0d6}, and recent work attempts to learn
example selectors \cite{wang2024learning}.  
In information retrieval (IR), most existing LLM-based QE pipelines rely on hand-crafted prompts or
manually curated exemplars drawn from mismatched domains
\cite{wang2023query2doc}, often yielding unstable performance.  
Moreover, nearly all prior studies use a single LLM, leaving unexplored how
to combine the complementary knowledge of multiple LLMs in a 
training-free manner.

This paper proposes a fully automated, domain-adaptive framework for
ICL-based query expansion that eliminates manual prompt design and enables
multi-LLM expansion.  
Given an unlabeled target corpus, we first construct a large in-domain
example pool via a BM25$\rightarrow$MonoT5 pipeline \cite{nogueira2020document}.
We then select diverse few-shot exemplars using a simple clustering
strategy over Contriever embeddings.  
This produces stable and domain-matched demonstrations without human
annotation.

Beyond single-LLM QE, we introduce a {two-LLM expansion ensemble}:
two heterogeneous LLMs independently generate expansions using the same
cluster exemplars, and a refinement LLM synthesizes them into one coherent,
noise-reduced expansion.  
This query-level fusion exploits complementary lexical and semantic cues
from both models without additional training or multiple retrieval runs.

Using Qwen-2.5-7B-Instruct as the generator and refinement model, we
evaluate on DL20, DBPedia-Entity, and SciFact.
Cluster-based exemplars consistently outperform BM25, Rocchio, and
hand-crafted few-shot baselines, while our refined two-LLM ensemble yields
the strongest gains
and many statistically significant.
Our contributions are:
\begin{itemize}
    \item A fully automated, label-free pipeline that builds large in-domain QE exemplar pools via BM25–MonoT5.
    \item A simple and reproducible clustering strategy for selecting stable, diverse ICL demonstrations.
    \item A training-free two-LLM expansion ensemble with LLM refinement, producing robust improvements across domains.
\end{itemize}

\section{Related Work}
\label{sec:related}

\smallsection{Query Expansion in IR}
Classical QE relies on PRF,
with Rocchio \cite{rocchio1971relevance} and RM3 \cite{abdul2004umass}
representing canonical approaches.
These methods often suffer from query drift when feedback documents are
noisy \cite{cao2008selecting,miao2012proximity,liu2022simple}.
Neural QE methods such as doc2query/doc2query-T5 \cite{nogueira2019document}
generate pseudo-queries offline, while recent LLM-based QE uses prompting
\cite{jagerman2023query,wang2023query2doc} to produce semantically richer
expansions.
However, unconstrained LLM generation may hallucinate or introduce
off-domain terms, motivating corpus-grounded approaches such as CSQE
\cite{lei2024corpus} or MUGI \cite{zhang2024exploring}, which require
multi-stage retrieval or additional reranking.

Our work differs by constructing {in-domain} pseudo-relevant
examples offline, then using them as demonstrations for training-free
few-shot ICL-based QE, avoiding iterative retrieval or human-written
examples.

\smallsection{In-Context Learning and Example Selection}
ICL is sensitive to the choice and order of examples 
\cite{NEURIPS2020_1457c0d6}.
Example selection has been explored via semantic similarity
\cite{liu2021makes}, diversity-based voting \cite{su2022selective},
information-theoretic metrics
\cite{sorensen2022information,gonen2022demystifying,li2023finding},
and supervised selectors \cite{wang2024learning}.
In IR, systematic exemplar selection remains underexplored.
We contribute a simple, training-free clustering approach that produces
diverse, domain-aligned examples from a pseudo-labeled pool.

\smallsection{Multi-Agent and Multi-LLM Collaboration}
Recent work views LLMs as collaborative agents, where multiple
specialized or redundant models communicate, critique, or refine each
other's outputs
\cite{dong2024self,guo2024large}.
Multi-agent debate \cite{liang2024encouraging} and self-consistency refinement
\cite{wangself} show that combining diverse model
perspectives can improve robustness and factuality.
Other systems adopt {LLM-as-a-judge} refiners 
\cite{chen2024mllm} to merge or validate multi-LLM outputs.

Our method is aligned with this direction but differs in its {retrieval} focus:
two heterogeneous LLMs independently generate candidate expansions, and a
third refinement LLM consolidates them into a single coherent expansion
{before retrieval}.
This produces a training-free, query-level fusion mechanism that avoids
multiple retrieval passes and leverages complementary LLM knowledge for QE.

\section{Methodology}
\label{sec:method}

\begin{figure}[t]
  \centering
  \includegraphics[width=0.85\columnwidth]{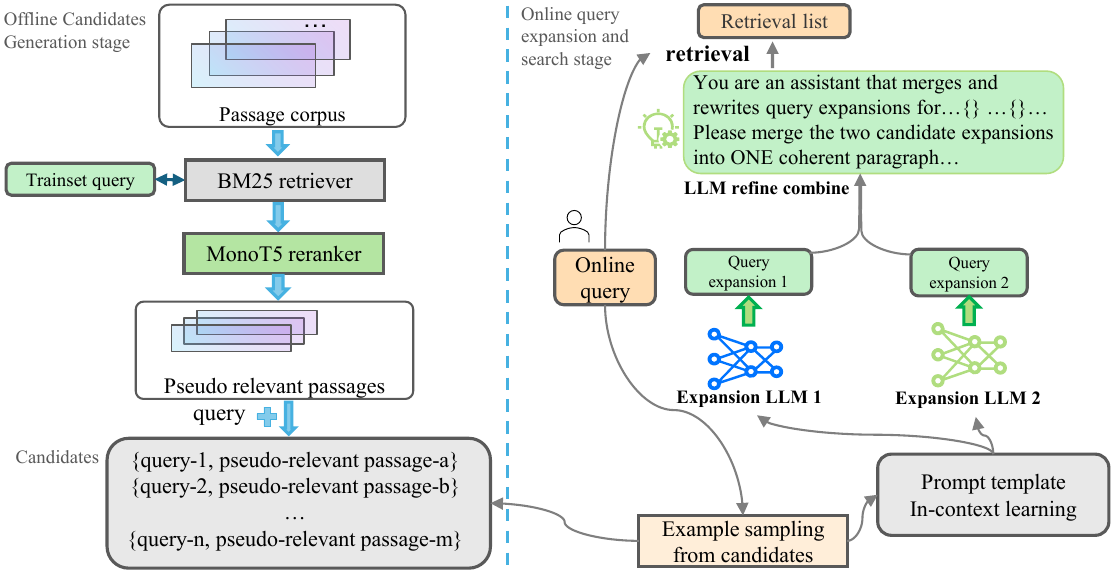}
  \caption{Overview of our automated pipeline for constructing
  domain-adaptive few-shot candidate pools, selecting cluster-based
  demonstrations, and performing two-LLM query expansion with LLM
  refinement.}
  \label{fig:archi}
\end{figure}

Our goal is to automatically construct an in-domain example pool for
few-shot query expansion, and then leverage LLMs to generate
high-quality expansions that improve retrieval effectiveness.
Our framework which is shown in Fig.~\ref{fig:archi}, consists of three stages:
(1) automatic pseudo-relevance pool construction,
(2) few-shot expansion generation with exemplar selection,
and (3) multi-LLM expansion ensemble with LLM refinement.

\subsection{Stage 1: Automatic In-Domain Example Pool}
\label{sec:example-generation}

Given a domain corpus, we build a pool $\mathcal{C}$ of
(query, expansion) pairs without human labels.
Seed queries (training queries of each dataset) are used only to retrieve
content.  
For each query, we obtain the BM25 top-$N$ candidates, rerank them using
MonoT5, and take the top-1 passage as a pseudo-relevant expansion.
This yields hundreds or thousands in-domain examples (e.g., 100k for MS MARCO, 809 for SciFact).
These serve as the few-shot ICL pool for Stage~2.

\subsection{Stage 2: Few-Shot Query Expansion with LLMs}
\label{sec:fewshot-qe}

For each test query, we construct a conversational prompt consisting of:
a system instruction, $k$ demonstration pairs from $\mathcal{C}$, and the
test query.  
We adopt the Qwen2.5-7B-Instruct model and generate a maximum 64 tokens expansion,
which is concatenated with the 5 repeated original query following~\cite{wang2023query2doc}.

\paragraph{Demonstration selection.}
We use a single training-free strategy: \textbf{Cluster}.
Contriever-based embeddings are computed for all candidates; $k$-means
clusters the pool into $k$ semantic groups.
We select the medoid of each cluster as the exemplar:
\[
\mathcal{E}(q)=\{\operatorname{medoid}(\mathcal{C}_1),\dots,
\operatorname{medoid}(\mathcal{C}_k)\}.
\]

Algorithm~\ref{alg:cluster} presents the clustering-based strategy, which partitions the candidate pool into $k$ semantic clusters and selects the medoid (center-nearest) example from each cluster to ensure topical diversity. This produces a diverse and stable demonstration set.

\begin{algorithm}[htbp]
\caption{Cluster-based Selection of $k$ Representative Examples}
\label{alg:cluster}
\KwIn{
    Example pool $\mathcal{E} = \{(q_i, p_i)\}$; \\
    embedding model $f(\cdot)$, contriever; \\
    number of clusters $k$
}
\KwOut{Cluster-diverse subset $\mathcal{E}_k$}

\BlankLine
\ForEach{$(q_i, p_i) \in \mathcal{E}$}{
    Encode $\mathbf{v}_i \leftarrow f(q_i + p_i)$ \tcp*{Joint query-passage embedding}
}
Normalize all $\mathbf{v}_i$ to unit length\;

Run $k$-means clustering on $\{\mathbf{v}_i\}$, yielding centroids $\{\mathbf{c}_1, \dots, \mathbf{c}_k\}$\;

\For{$j = 1$ \KwTo $k$}{
    Let $C_j$ be the set of examples in the $j$-th cluster\;
    Select $(q^*, p^*) \in C_j$ closest to $\mathbf{c}_j$ in $\ell_2$ distance\;
    Add $(q^*, p^*)$ to $\mathcal{E}_k$
}

\Return{$\mathcal{E}_k$}
\end{algorithm}

\subsection{Stage 3: Two-LLM Expansion Ensemble with LLM Refinement}
\label{sec:llmEnsembleRefine}

To exploit inter-model diversity, we expand each query using two
independently prompted LLMs,
$\mathrm{LLM}^{(1)}$ and $\mathrm{LLM}^{(2)}$, obtaining
$e^{(1)}(q)$ and $e^{(2)}(q)$.
Instead of score-level fusion, we introduce a third LLM
(Qwen2.5-7B) as a refinement module that consolidates the two expansions
into one coherent and richer expansion:
\[
\tilde{e}(q)
= f_{\mathrm{ref}}\bigl(q, e^{(1)}(q), e^{(2)}(q)\bigr).
\]
The refinement instruction asks the model to keep useful entities,
relations, and domain knowledge from both expansions, while eliminating
redundancy and noise.  
We limit $\tilde{e}(q)$ to at most 128 tokens during generation.

Following \cite{wang2023query2doc}, 
the final expanded query concatenates five copies of the original query with the refined expansion:
\[
\hat{q} = \underbrace{q \oplus q \oplus q \oplus q \oplus q}_{\text{5 copies of } q} \oplus \tilde{e}(q).
\]
which is then used for retrieval.

\subsection{Prompts}
\label{sec:prompts}

\begin{figure}[htbp]
    \centering
    \includegraphics[width=0.95\linewidth]{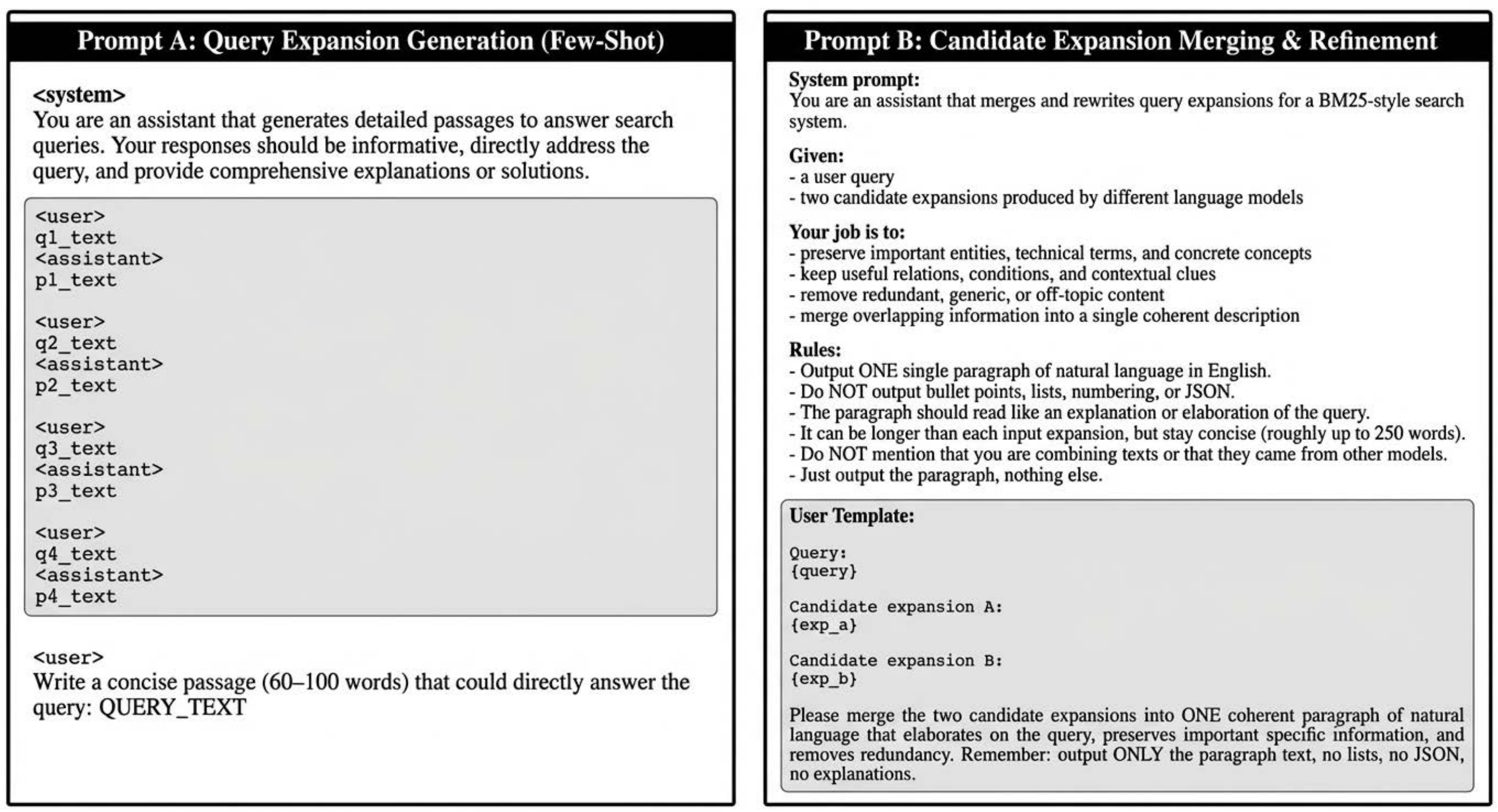}
    \caption{Illustration of the two prompts used in our framework:
    (i) the expansion-generation prompt containing system
    instruction, four exemplar query–passage pairs, and the test query;
    and (ii) the refinement prompt that consolidates two candidate
    expansions into one coherent expansion.}
    \label{fig:prompts}
\end{figure}

To illustrate our prompting strategy, we show a compact diagram covering both stages of our pipeline.
The expansion-generation prompt provides the LLM with a system instruction, four in-domain query–passage exemplars, and the test query.
The refinement prompt then takes the original query together with two independently generated expansions, and a second LLM produces a single consolidated expansion.
Fig.~\ref{fig:prompts} summarizes both prompts in a unified box-style chat diagram.

\section{Experimental Setup}
\label{sec:experimental-setup}

\subsection{Tasks and Corpora}
\paragraph{Datasets}
We evaluate our methods on three public retrieval benchmarks spanning web search, entity-oriented search, and scientific claim verification:

\begin{itemize}
  \item \textbf{TREC DL20} \cite{craswell2021overviewtrec2020deep}:  
        54 queries with graded relevance judgments, evaluated on the MS MARCO passage corpus (8.8M passages).  
        This benchmark represents large-scale, high-recall web search.

  \item \textbf{DBPedia-Entity} \cite{thakur2beir}:  
        400 entity-centric queries over 4.9M Wikipedia abstracts, focusing on entity grounding and fine-grained semantics.

  \item \textbf{SciFact} \cite{wadden-etal-2020-fact}:  
        300 scientific claims paired with 5k biomedical abstracts, annotated for {support} or {refute}.  
        This dataset stresses precise evidence retrieval under a compact corpus.

\end{itemize}

Datasets statistic is shown in Table~\ref{tab:data-stats}.
Across these datasets, corpus sizes and topic domains vary  from millions of heterogeneous web documents to a small specialized biomedical collection (SciFact), allowing us to evaluate the robustness of multi-LLM query expansion.

\begin{table}[ht]
\centering
\small
\scalebox{0.90}{
\begin{tabular}{lrrr}
\toprule
Dataset & \#Test Queries & Corpus Size & Generated Pool Size \\
\midrule
DL20       & 54   & 8.8 millions   & 100000 \\
DBPedia    & 400  & 4.9 millions   & 67 \\
SciFact    & 300  & 5000     & 809 \\
\bottomrule
\end{tabular}}
\caption{Evaluation datasets and statistics.}
\label{tab:data-stats}
\end{table}

\subsection{Indexing, Pseudo-Relevance Harvesting, and LLM Expansion}

All corpora are indexed using the Anserini Java 11 toolkit
\cite{yang2018anserini} with BM25 (default $k_{1}{=}0.9$, $b{=}0.4$) as the retriever. During online evaluation, BM25 retrieves the top-100
passages for each test query.

For constructing the in-domain exemplar pools, we apply the same BM25
retrieval offline and rerank the top-100 candidates using MonoT5-3B
\cite{nogueira2020document}. The highest-scoring passage per query is kept,
yielding three pseudo-relevance pools (Table~\ref{tab:data-stats}):  
100k passages for DL20 (MS MARCO), 67 for DBPedia-Entity (from its dev set),
and 809 for SciFact (from its training set). These passage–query pairs serve
as the candidate demonstrations for few-shot prompting.

All query expansions except query expansion 2 (in Fig.~\ref{fig:archi}, the Expansion LLM2) and the LLM used to refine two expansions are generated by Qwen-2.5-7B-Instruct
\footnote{\url{https://huggingface.co/Qwen/Qwen2.5-7B-Instruct}}
without fine-tuning. The Expansion LLM2 is Llama-3.1-8B-Instruct\footnote{\url{https://huggingface.co/meta-llama/Llama-3.1-8B-Instruct}}. Each prompt includes the task system message, four
\((q_i, p_i)\) exemplars (each passage truncated to 60 words), and the test
query, fitting within a 1024-token window.
Expansions are produced using a 4-beam decoding strategy. The generation is capped at 64 new tokens (for LLM refine module, it is 128), and repetition is controlled through a penalty term together with a constraint that prevents the model from generating repeated 2-gram sequences.
All experiments run on a single Nvidia
A100 GPU; generating expansions for several hundred queries typically
finishes within minutes.

\subsection{Evaluation Metrics and Baselines}
\label{sec:metrics-baselines}

\paragraph{Metrics}
We report NDCG@10, P@10, and Recall@100 using
trec\_eval\footnote{\url{https://github.com/usnistgov/trec\_eval}}.
Higher values indicate better retrieval effectiveness.

\paragraph{Baselines}
We compare against standard lexical and LLM-based query expansion methods:

\begin{itemize}
  \item {BM25}: Retrieval without expansion.

  \item {BM25+Rocchio}: Classical PRF using
        Anserini’s default configuration.

  \item {ZeroShot}: LLM-generated expansion using only the prompt without demonstrations.

  \item {FewShot4-Fixed}: Few-shot prompting with the fixed exemplars
        from \cite{wang2023query2doc}, shared across all queries.
\end{itemize}

\paragraph{Our approaches}
All proposed models use identical corpora, Lucene indexes, and decoding
settings; differences arise solely from how expansion is produced.

\begin{itemize}
  \item {Cluster-ICL QE}: Four in-domain exemplars selected via
        $k$-means clustering over Contriever embeddings.

  \item {Two-LLM QE (Concat)}: Expansions from two LLMs concatenated
        into a single expanded query.

  \item {Two-LLM QE (Refine)}: Two expansions merged by an additional
        LLM that rewrites them into a coherent, unified expansion.
\end{itemize}

This setup isolates the effect of exemplar selection and multi-LLM
combination, enabling controlled comparison across all variants.

\section{Results and Analysis}
\label{sec:results}

\subsection{Main Results}

\begin{table*}[htbp]
\centering
\small
\setlength{\tabcolsep}{3pt}
\scalebox{0.90}{
\begin{tabular}{lcccccccccc}
\toprule
& \multicolumn{3}{c}{\textbf{DL20}} 
& \multicolumn{3}{c}{\textbf{DBPedia}} 
& \multicolumn{3}{c}{\textbf{SciFact}} \\
\cmidrule(lr){2-4} \cmidrule(lr){5-7} \cmidrule(lr){8-10}
\textbf{Method} 
& \textbf{N10} & \textbf{P10} & \textbf{R100} 
& \textbf{N10} & \textbf{P10} & \textbf{R100}
& \textbf{N10} & \textbf{P10} & \textbf{R100} \\
\midrule
BM25 
& 47.96$^*$ & 53.89$^*$ & 48.34$^*$ 
& 31.80$^*$ & 28.20$^*$ & 46.82$^*$ 
& 67.89$^*$ & 8.83$^*$ & 92.53$^*$ \\

BM25+Rocchio 
& 49.10$^*$ & 57.22$^*$ & 52.34$^*$ 
& 30.76$^*$ & 28.58$^*$ & 47.11$^*$ 
& 62.78$^*$ & 8.80$^*$ & 91.87$^*$ \\
\midrule

ZeroShot 
& 53.49$^*$ & 59.44$^*$ & 51.00$^*$ 
& 35.78$^*$ & 30.17$^*$ & 51.70$^*$ 
& 68.90$^*$ & 9.07$^*$ & 94.70$^*$ \\

FewShot4-Fixed 
& 56.38$^*$ & 64.44$^*$ & 55.72$^*$ 
& 36.54$^*$ & 30.52$^*$ & 50.68$^*$ 
& 69.19$^*$ & 9.10$^*$ & 94.37$^*$ \\
\midrule

Cluster-ICL QE 
& 58.71$^*$ & 63.70$^*$ & 56.68 
& 36.89$^*$ & 31.10$^*$ & 52.38$^*$ 
& 69.69$^*$ & 9.20$^*$ & 94.00$^*$ \\
\midrule

Two-LLM QE (Concat) 
& 59.47$^*$ & 64.81$^*$ & 58.22 
& 38.67 & 31.47 & 52.74$^*$ 
& 71.86 & 9.47 & 94.60$^*$ \\

Two-LLM QE (Refine) 
& \overallbest{62.86} & \overallbest{68.33} & \overallbest{58.51}
& \overallbest{39.14} & \overallbest{32.15} & \overallbest{53.64}
& \overallbest{72.07} & \overallbest{9.50} & \overallbest{95.60} \\
\bottomrule
\end{tabular}}
\caption{
Retrieval performance (\%).  
N10 = NDCG@10, P10 = P@10, R100 = Recall@100.  
$^*$ means that Refine method is significantly better than
the corresponding method (paired t-test, $p\le0.05$).  
Shaded bold cells denote the overall best per metric.
}
\label{tab:combined-results}
\end{table*}

\smallsection{Lexical baselines}
As shown in Table~\ref{tab:combined-results}, BM25 and BM25+Rocchio form 
competitive lexical baselines, but their effectiveness varies by domain:
Rocchio improves deep recall on DL20 but often reduces early precision on
DBPedia and SciFact due to query drift. These results highlight the 
limitations of corpus-only term expansion compared to LLM-generated rewriting.

\smallsection{Zero-shot vs.\ fixed exemplars}
ZeroShot consistently outperforms lexical baselines, demonstrating that 
Qwen-2.5-7B-Instruct can generate useful paraphrases without demonstrations.
However, FewShot4-Fixed yields only small, domain-sensitive changes: it modestly 
helps when in-domain (DL20) but offers limited or negative gains on DBPedia and 
SciFact, confirming the weakness of out-of-domain exemplars.

\smallsection{Impact of in-domain cluster exemplars}
Cluster-ICL QE delivers clear and robust improvements over both ZeroShot and
FewShot4-Fixed, 
e.g., on SciFact it improves NDCG@10 from 69.19 (FewShot4-Fixed) to 69.69 (Table \ref{tab:combined-results}).
Its demonstrations, harvested via BM25→MonoT5 pseudo labelling, better match domain terminology and yield consistent gains across all datasets despite varying pool sizes. This validates the importance of domain-adaptive 
few-shot construction.

\smallsection{Multi-LLM expansion and refinement}
Direct concatenation of expansions from two LLMs (Concat) provides
small but consistent gains over single-model Cluster-ICL QE.  
The refinement variant (Refine) achieves the best results on all 
datasets, with +4.15, +2.25, and +2.38 absolute gains on NDCG@10 over Cluster-ICL QE on DL20, DBPedia, and SciFact, respectively.  
Refinement removes redundancy and synthesizes complementary cues from both models, 
acting as a training-free query-level fusion mechanism.

Overall, the results demonstrate that (i) in-domain exemplar construction is more
effective than fixed few-shot prompting, and (ii) multi-LLM refinement provides the
largest and most stable improvements across datasets.

\subsection{Extension to Dense Retrieval on TREC DL20}
\label{sec:dense-dl20}

Thus far, all results have used BM25 as the first-stage retriever.
To verify that our query expansion strategies are not tied to lexical
retrieval, we further evaluate on {TREC DL20} with a dense
encoder-based retriever.
Specifically, we reuse exactly the same expanded queries as in the BM25
experiments, but replace BM25 with a Sentence-BERT retriever implemented in
Pyserini.
We use the pre-built FAISS index
\texttt{msmarco-v1-passage.sbert} together with the
\texttt{sentence-transformers/msmarco-distilbert-base-v3} encoder.

Table~\ref{tab:dl20-sbert-results} reports dense-retrieval effectiveness.
The SBERT baseline (no expansion) already outperforms BM25 on NDCG@10 and
P@10, but still lags behind our expansion-based methods.
Cluster-ICL QE improves over SBERT by +2.4 NDCG@10 (63.44 $\rightarrow$
65.84) while maintaining comparable recall.
Two-LLM QE (Concat) and Two-LLM QE (Refine) deliver the strongest gains,
reaching 68.59/68.32 NDCG@10 and up to 71.67 P@10, with modest improvements
in Recall@100.
These results indicate that our multi-LLM expansion and refinement pipeline
transfers beyond BM25 and can also serve as a plug-in enhancement for
dense retrievers such as SBERT.

\begin{table*}[htbp]
\centering
\scalebox{1.0}{
\begin{tabular}{lccc}
\toprule
\textbf{Method} & \textbf{NDCG@10} & \textbf{P@10} &  \textbf{Recall@100} \\
\midrule
SBERT                     & 63.44$^*$ & 68.33$^*$  & 50.89$^*$ \\
\midrule
ZeroShot                  & 56.89$^*$ & 58.52$^*$  & 46.15$^*$ \\
FewShot4‑Fixed           & 62.79$^*$ & 65.37$^*$  & 48.37$^*$ \\
\midrule
\multicolumn{4}{l}{\textbf{Proposed approaches}}\\
\midrule
Cluster-ICL QE          & 65.84$^*$ & 68.33$^*$  & 49.39$^*$ \\
\midrule
Two-LLM QE (Concat) & \overallbest{68.59} & 70.74  & 51.45 \\
Two-LLM QE (Refine) & 68.32 & \overallbest{71.67}  & \overallbest{52.29} \\
\bottomrule
\end{tabular}}
\caption{Dense retrieval (SBERT) performance (\%) on {TREC DL20}. $^*$ means Refine method is significantly better than
the corresponding method (paired t-test, $p\le0.05$). }
\label{tab:dl20-sbert-results}
\end{table*}

\subsection{Ablation Study}

\subsubsection{Two-LLM Ensembles for Fixed vs.\ Cluster-Selected Demonstrations}

\begin{figure}[tb]
  \centering

  \subfigure[NDCG@10 on TREC DL20]{
    \includegraphics[width=0.48\linewidth]{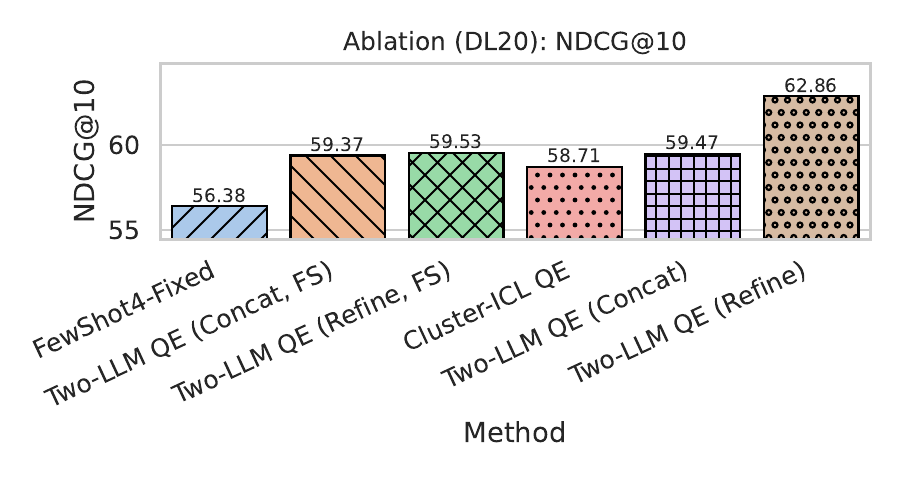}
  }\hfill
  \subfigure[Recall@100 on TREC DL20]{
    \includegraphics[width=0.48\linewidth]{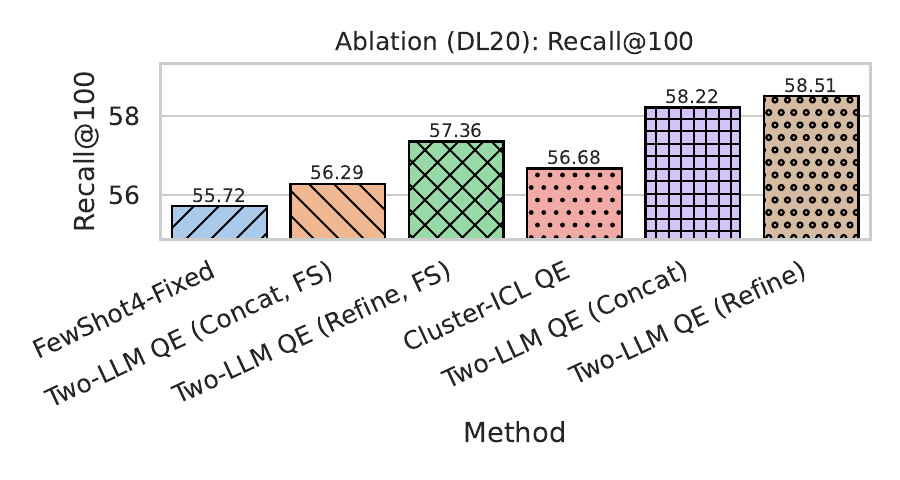}
  }

  \caption{Ablation on TREC DL20: comparison of FewShot4-Fixed vs.\ in-domain Cluster-ICL QE under two-LLM ensembles.}
  \label{fig:dl20-ablation-ensemble}
\end{figure}

Figure~\ref{fig:dl20-ablation-ensemble} compares two-LLM ensembling on
(i) the fixed FewShot4-Fixed exemplars and (ii) our in-domain Cluster-ICL QE.
For FewShot4-Fixed, adding a second LLM yields only modest gains in NDCG@10
and Recall@100, suggesting that model diversity can partly offset
suboptimal, out-of-domain exemplars.
In contrast, ensembling on top of Cluster-ICL QE brings substantially larger
improvements, with the refinement-based ensemble achieving the best scores.
This shows that high-quality in-domain exemplars and multi-LLM refinement
are complementary, jointly leading to stronger retrieval effectiveness.

\subsubsection{LLM1 vs.\ LLM2 vs.\ Two-LLM Ensemble Performance}
\label{sec:ablation-llm1-llm2}

\begin{figure}[tb]
  \centering

  \subfigure[NDCG@10 on TREC DL20]{
      \includegraphics[width=0.48\linewidth]{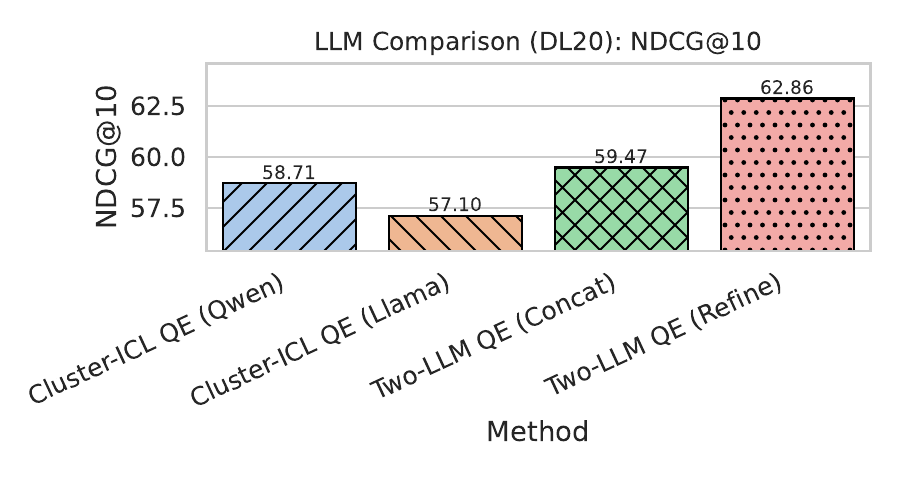}
  }\hfill
  \subfigure[Recall@100 on TREC DL20]{
      \includegraphics[width=0.48\linewidth]{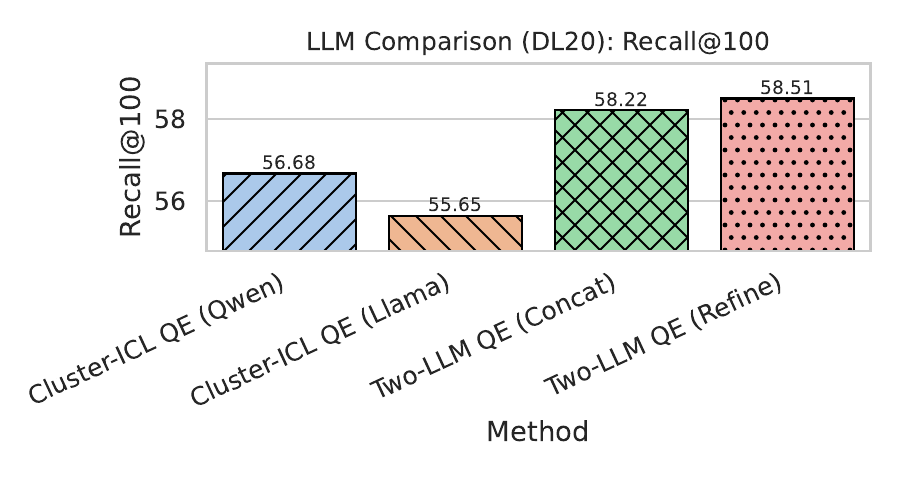}
  }

  \caption{Comparison of single-LLM vs.\ two-LLM ensemble under the Cluster-ICL QE setting.  
  Ensemble-Refine provides the strongest performance across both early-precision (NDCG@10) and deep-recall metrics.}
  \label{fig:dl20-llm-ablation}
\end{figure}

Figure~\ref{fig:dl20-llm-ablation} compares two single-LLM expansions (Qwen, LLaMA) with their corresponding direct and refinement ensembles under the Cluster-ICL setting.

Qwen already outperforms LLaMA across all metrics, but both single-model variants trail behind the two ensemble methods.
Direct concatenation yields small but consistent gains, indicating that the two LLMs provide complementary expansion cues.
The refinement ensemble achieves the strongest performance overall, improving NDCG@10 by +4.1 and Recall@100 by +1.8 over Qwen.

These results show that (i) heterogeneous LLMs encode different but useful expansion behaviours, and (ii) an LLM-based refinement step integrates these signals more effectively than simple text concatenation.
Thus, multi-LLM refinement is a compact, training-free, and principled way to boost ICL-based query expansion.

\subsubsection{Effect of Expansion Length: Longer Single Expansion vs. Two-LLM Refined QE}

To understand whether simply generating longer LLM expansions improves retrieval, we compare our default max 64 token setting with an extended experiment with max 128 token variant.
Simply increasing the maximum expansion length from 64 to 128 tokens degrades effectiveness on DL20 (e.g., NDCG@10 drops from 58.71 to 54.26), suggesting that overly long generations introduce noise and topic drift. In contrast, Two-LLM QE (Refine) reaches 62.86 NDCG@10 and 68.33 P@10, indicating that quality-controlled multi-LLM refinement, rather than verbosity, is the key driver of QE gains.

\section{Conclusion}
\label{sec:conclusion}

We presented a fully automated and label-free framework for in-context query expansion with LLMs.
Our method constructs large pseudo-relevant in-domain exemplar pools from an unlabeled corpus, selects diverse demonstrations via simple clustering, and enables robust, training-free few-shot prompting.
To further exploit complementary generation patterns, we introduced a two-LLM expansion ensemble with an additional refinement LLM that synthesizes multiple expansions into a single coherent query-side augmentation.

Across TREC DL20, DBPedia, and SciFact, our cluster-based exemplars consistently outperform zero-shot and fixed few-shot baselines, while the refined two-LLM ensemble yields the best overall gains, with many of these improvements being statistically significant.
Our results highlight the importance of domain-matched exemplars and cross-LLM refinement for stable, high-quality QE.
All candidate pools and code are released to support future research on scalable, domain-adaptive LLM-based retrieval.

%
%
%
\bibliographystyle{splncs04}
\bibliography{mybibliography}
%




\end{document}